\documentclass[aps, prd, amsmath, floats, floatfix, twocolumn, nofootinbib, showpacs]{revtex4-1}
\usepackage{graphicx}
\usepackage{color}
\usepackage{latexsym}

\newcommand{\bi}{\bibitem}
\newcommand{\be}{\begin{eqnarray}}
\newcommand{\ee}{\end{eqnarray}}
\newcommand{\rar}{\rightarrow}

\begin{document}

\title{Can static regular black holes form from gravitational collapse?}

\author{Yiyang Zhang}

\author{Yiwei Zhu}

\author{Leonardo Modesto}

\author{Cosimo Bambi}
\email[Corresponding author: ]{bambi@fudan.edu.cn}

\affiliation{Center for Field Theory and Particle Physics \& Department of Physics, Fudan University, 200433 Shanghai, China}

\date{\today}

\begin{abstract}
Starting from the Oppenheimer-Snyder model, we know how in classical general 
relativity the gravitational collapse of matter form a black hole with a central 
spacetime singularity. It is widely believed that the singularity must be removed 
by quantum gravity effects. Some static quantum-inspired singularity-free black hole 
solutions have been proposed in the literature, but when one considers simple 
examples of gravitational collapse the classical singularity is replaced by a bounce, 
after which the collapsing matter expands for ever. We may expect 
3 possible explanations: $i)$ the static regular black hole solutions are not physical, 
in the sense that they cannot be realized in Nature, $ii)$ the final product of the 
collapse is not unique, but it depends on the initial conditions, or $iii)$ boundary 
effects play an important role and our simple models miss important physics. In 
the latter case, after proper adjustment, the bouncing solution would approach 
the static one. We argue that the ``correct answer" may be related to the appearance 
of a ghost state in de~Sitter spacetimes with super Planckian mass. Our black 
holes have indeed a de~Sitter core and the ghost would make these configurations 
unstable. Therefore we believe that these black hole static solutions represent 
the transient phase of a gravitational collapse, but never survive as 
asymptotic states.
\end{abstract}

\maketitle


\section{Introduction}

In classical general relativity, under the main assumptions of the validity of the 
strong energy condition and of the existence of global hyperbolicity, the collapse
of matter inevitable produces a singularity of the spacetime~\cite{sing}. At a 
singularity, predictability is lost and standard physics breaks down. According to 
the weak cosmic censorship conjecture, spacetime singularities formed from 
collapse must be hidden behind an event horizon and the final product of the collapse 
must be a black hole~\cite{wccc}. In 4-dimensional general relativity, the only 
uncharged black hole solution is the Kerr metric~\cite{hair}, which reduces to the 
Schwarzschild solution in the spherically symmetric case. The Oppenheimer-Snyder 
model is the simplest fully analytic example of gravitational collapse, describing 
the contraction of a homogeneous spherically symmetric cloud of dust~\cite{os}. 
It clearly shows how the collapse produces a spacetime singularity and the final 
product is a Schwarzschild black hole.

In analogy with the appearance of divergent quantities in other classical theories, 
it is widely believed that spacetime singularities are a symptom of the limitations 
of classical general relativity, to be removed by quantum gravity effects. While we 
do not have yet any robust and reliable theory of quantum gravity, the resolution 
of spacetime singularities has been investigated in many quantum gravity inspired 
models. Very different approaches have studied corrections to the Schwarzschild/Kerr
solution, finding black hole metrics in which the curvature invariants are 
always finite~\cite{regular}\footnote{We note that it may also be
possible that the quantum corrections that smooth out the singularity may be 
intrinsically-quantum and not reducible to the metric form. In such a case, the metric 
description would simply break down.}. In the same spirit, one can study the modifications 
to the Oppenheimer-Snyder solution and to other models of collapse. In this case, 
the singularity is replaced by a bounce, after which the cloud starts expanding~\cite{bounce}. 
It is therefore disappointing that the quantum-gravity corrected model of collapse 
does not reproduce to the quantum-gravity corrected black hole solution.

In this paper, we want to investigate this apparent contradictory result. First, we 
determine both the quantum-gravity corrected static black hole metric and the 
quantum-gravity corrected homogeneous collapse solution within the same 
theoretical framework, since the ones reported in the literature come from 
different models. We find that the problem indeed exists. Second, we try to figure
out the possible reason. One
possibility is that the static regular black hole spacetimes are {\it ad hoc} solutions, 
but they cannot be created in a collapse and therefore they are physically irrelevant. 
The collapse always produces an object that bounces. Another possible explanation is 
that the final product of the collapse depends on the initial conditions. The collapse 
of a homogeneous cloud creates an object that bounces, while with other initial 
conditions (not known at present) the final product is a static regular black hole.
Lastly, it is possible that the simple homogeneous collapse oversimplifies the 
model, ingoing and outgoing energy fluxes between the interior and the exterior 
solutions are important, and, after proper readjustment that seems to be difficult to 
have under control within an analytic approach, the collapsing model approaches 
the static regular black hole solution. Our quantum-gravity inspired theories 
are unitary, super-renormalizable or finite at quantum level, and there are no 
extra degrees of freedom at perturbative level around flat spacetime. This should 
rule out the possibility that the explanation of our puzzle is due to the fact that 
these models may not be consistent descriptions of quantum gravity.
However, these theories display a ghost state  in de~Sitter spacetime 
when the cosmological constant exceeds the square of the Planck mass. This fact may
be responsible for our finding and answers the question in the title of this paper. 
Our black holes have indeed a de~Sitter core with
an effective cosmological constant larger than the square of the Planck mass
when the black hole mass exceeds the Planck mass. The presence of a ghost
makes the solutions unstable and therefore they cannot be the final 
product of the gravitational collapse.

The content of the paper is as follows. In the next section, we briefly review the
classical homogeneous and spherically symmetric collapse model. In 
Section~\ref{s-3}, we derive the spherically 
symmetric black hole solutions in a super-renormalizable and asymptotically-free 
theory of gravity with the family of form factors proposed by Krasnikov~\cite{ff-k} and 
Tomboulis~\cite{ff-t}. In Section~\ref{s-4}, we study the spherically symmetric 
homogeneous collapse in the same models. Summary and conclusions are in 
Section~\ref{s-5}.

\section{Black holes and gravitational collapse in classical general relativity \label{s-2}}

The most general spherically symmetric metric describing a collapsing cloud of 
matter in comoving coordinates is given by
\be
ds^2=-e^{2\nu}dt^2+\frac{R'^2}{G}dr^2+R^2d\Omega^2 \; ,
\ee
where $d\Omega^2$ represents the line element on the unit two-sphere 
and $\nu$, $R$, and $G$ are functions of $t$ and $r$. The energy-momentum 
tensor is given by 
\be
T^\mu_\nu={\rm diag}( -\rho(r,t), p_r(r,t),  p_\theta(r,t), p_\theta(r,t) ) \; ,
\ee
and the Einstein equations reduces to
\be
\frac{\kappa^2}{2} \rho &=& \frac{F'}{R^2R'}  \; , \label{eq-rho} \\
\frac{\kappa^2}{2} p_r &=& - \frac{\dot{F}}{R^2\dot{R}}\; , \label{eq-p} \\ 
\nu' &=& 2 \frac{p_\theta - p_r}{\rho + p_r} \frac{R'}{R} -  \frac{p'_r}{\rho + p_r} \; , \label{eq-nu} \\ 
\dot{G} &=& 2\frac{\nu'}{R'}\dot{R}G \; , \label{eq-g}
\ee
where the $'$ denotes a derivative with respect to $r$, and the $\dot{}$ 
denotes a derivative with respect to $t$. The function $F(r,t)$ is called 
Misner-Sharp mass and is defined by
\be\label{misner}
F=R(1-G+e^{-2\nu}\dot{R}^2) \; .
\ee
The whole system has a gauge degree of freedom that can be fixed by 
setting the scale at a certain time. One usually sets the area radius $R(r,t)$ to
be equal to the comoving radius $r$ at the initial time $t_{\rm i}=0$, i.e.
$R(r,0)=r$. We can then introduce a scale factor $a$
\begin{equation}
R(r,t)=ra \; ,
\end{equation}
that will go from 1, at the initial time, to 0, at the time of the formation of the 
singularity. The condition to describe collapse is $\dot{a}<0$.

For a homogeneous perfect fluid, $p_r = p_\theta = p(t)$. The simplest case
is the gravitational collapse of a cloud of dust, $p=0$, which is the well-known 
Oppenheimer-Snyder model~\cite{os}. From Eq.~(\ref{eq-rho}), we see that $F$ 
is proportional to the amount of matter enclosed within the shell labeled by $r$ 
at the time $t$. For dust, from Eq.~(\ref{eq-p}) it follows that $F$ is independent 
of $t$, so there are no inflows and outflows through any spherically symmetric 
shell of radial coordinate $r$. If $r_{\rm b}$ denotes the comoving radial coordinate of 
the boundary of the cloud, $F(r_{\rm b})=2M_{\rm Sch}$, where $M_{\rm Sch}$ is the  
Schwarzschild mass of the vacuum exterior. Let us note that in the general case
of a perfect fluid that is not true, and for a non-vanishing pressure the homogeneous 
spherically symmetric interior must be matched with a non-vacuum Vaidya 
exterior spacetime. Eq.~(\ref{eq-nu}) reduces to $\nu' = 0$ and one can always choose 
the time gauge in such a way that $\nu=0$. From Eq.~(\ref{eq-g}), we find that $G$ 
is independent of $t$ and we can write $G=1+f(r)$. In the homogeneous marginally 
bound case (representing particles that fall from infinity with zero initial velocity), 
$f=0$ and therefore $G=1$.

In the homogeneous spherically symmetric gravitational collapse of a cloud of
dust, one finds that the energy density is given by
\be
\rho(t) = \frac{\rho_0}{a^3} \, ,
\ee
where $\rho_0$ is the energy density at the initial time $t_{\rm i} = 0$, and
the scale factor is
\be
a(t) = \left(1-\frac{\sqrt{3 \rho_0}}{2} t\right)^{2/3} \, . 
\ee
The model has a strong curvature singularity for $a = 0$, which occurs at the
time $t_{\rm s} = 2/\sqrt{3\rho_0}$, as can be seen from the divergence of the 
Kretschmann scalar
\begin{eqnarray}
R_{\mu\nu\rho\sigma} R^{\mu\nu\rho\sigma}
=12 \frac{\ddot{a}^2 a^2 + \dot{a}^4}{a^4} \; .
\end{eqnarray}
The boundary of the cloud collapses along the curve $R(r_{\rm b},t)=r_{\rm b} a(t)$ 
and the whole cloud becomes trapped inside the event horizon at the time 
$t_{\rm tr}<t_{\rm s}$ for which $R(r_{\rm b},t_{\rm tr})=2M_{\rm Sch}=r_{\rm b}^3 \rho_0/3$, 
so 
\begin{equation}
t_{\rm tr}=t_{\rm s} - \frac{4M_{\rm Sch}}{3} \, .
\end{equation}

For $p \neq 0$, the exterior solution is not Schwarzschild, but the Vaidya spacetime,
because there is a non-vanishing flux through the boundary $r_{\rm b}$ and 
$F(r_{\rm b},t)$ does depend on time. In the radiation case, one finds
\be
\rho(t) &=& \frac{\rho_0}{a^4} \, , \\
a(t) &=& \left(1-2\sqrt{\frac{\rho_0}{3}}t \right)^{1/2} \, ,
\ee  
where $\rho_0$ is the energy density at the initial time $t_{\rm i} = 0$ and the 
singularity occurs at the time $t_{\rm s}=\sqrt{3/\rho_0} / 2$. 
Like in the dust model, the final outcome is a Schwarzschild black hole.
For the generic case of a perfect fluid with equation of state $p = \omega \rho$,
the scale factor is (for $\omega \neq -1$)
\be
a(t) = \left(1 - \frac{t}{t_{\rm s}}\right)^{\frac{2}{3(\omega + 1)}} \, ,
\ee
where $t_{\rm s}$ is the time of the formation of the singularity
\be
t_{\rm s} = \frac{2}{(\omega + 1) \sqrt{3\rho_0}} \, .
\ee
The time of the formation of the event horizon is still given by
$R(r_b,t_{\rm tr})=F(r_b,t_{\rm tr})$ and occurs before the time $t_{\rm s}$.

\section{Quantum-gravity inspired black holes \label{s-3}}

We start from the classical Lagrangian of the renormalized theory 
in~\cite{lll},
\be\label{eq-theory}
\mathcal{L}=-\sqrt{-g} \Big\{ \frac{1}{\kappa^2} R  - G^{\mu\nu} 
\frac {V^{-1}(-\Box_\Lambda)-1}{\kappa^2 \Box_{\Lambda}}
R_{\mu\nu} \Big\} \, ,
\ee
where $\Box_\Lambda= p^\mu p_\mu /\Lambda^2$ and we use the 
signature $(-,+,+,+)$. The main properties of our theoretical framework are discussed 
in Appendix~\ref{appendix-0}, where we show that the theories studied in this 
paper are unitary, super-renormalizable or finite at the quantum level, and there 
are no extra degrees of freedom (ghosts or tachyons) 
in flat spacetime. We note that we are not considering ``local higher derivative
theories of gravity'', but ``weakly non-local theories of gravity''.
Here with non-locality we mean that we have an operator with an infinite number of derivatives, while in a local theory the number of these derivatives would be finite. Weakly because it is only the whole sum that makes the theory non-local. However, the non-locality is not enough to have a good theory. We need that the propagator is the standard one times an entire function without zeros, singularities or poles in the whole complex plane. In this case, the theory does not have ghosts by construction, because the residue of the propagator at the pole is the same as the one of general relativity. 
The regularization of the solutions is thus 
due to the choice of the form factor and to the absence of interactions 
at high energy or, in other words, to asymptotic freedom.
More details can be found in the last paper in~\cite{bounce}.
The equations of motion for the theory up to terms quadratic
in the curvature are 
\begin{equation}\label{modifyGR}
G_{\mu\nu}+O(R^2_{\mu\nu}) = \frac{\kappa^2}{2} V(-\Box_\Lambda) T_{\mu\nu} \, .
\end{equation}
The right hand side can be considered as an effective energy-momentum term, 
defined by $\mathcal{S}_{\mu\nu}=V(-\Box_\Lambda) T_{\mu\nu}$. Within this 
approximation, the left hand side is compatible with the Bianchi identity, so the 
effective energy-momentum tensor is conserved 
\be
\nabla^\mu\mathcal{S}_{\mu\nu}=
\nabla^\mu \left[ V(-\Box_\Lambda)T_{\mu\nu} \right]=0 \, .
\ee

Now we want to solve the field equations for a static and spherically symmetric 
source. From the static property, the four-velocity is $u^\mu=(u^0, \vec{0})$; that 
is, only the timelike component is non-zero, and $u^0=(-g_{00})^{-1/2}$. 
For simplicity, we consider a point source. The $T^0_0 $ component is given by 
$T^0_0=-M \frac{\delta(r)}{4 \pi r^2}=-M \delta^3(\vec{x})$. Spherical or 
Cartesian-like coordinates are adopted to make calculations easier. In the 
spherically symmetric case, the metric is assumed to have the Schwarzschild form,
\begin{equation}
ds^2=-F(r)dt^2 + \frac{dr^2}{F(r)} +r^2 d\Omega ^2 \;
\end{equation}
where $F(r)$ is
\begin{equation}\label{fr}
F(r)=1-\frac{2 m(r)}{r} \;
\end{equation}
and $m(r)$ is the mass enclosed within the radius $r$.

The effective energy-momentum tensor is defined by
\begin{equation}
\mathcal{S}^{\mu}_{\nu}\equiv V(-\Box_\Lambda) 
T^{\mu}_{\nu} ={\rm diag} ( -\rho^e, p_r^e, p_\theta^e, p_\theta^e ) \, .
\end{equation}
The $\mathcal{S}^0_0$ component can be rewritten as
\be\label{eff-rho}
\rho^e(\vec{x})&=&V(-\Box_\Lambda) M \delta^{(3)} (\vec{x}) \nonumber\\
&=&M \int \frac{\mathrm{d}^3 k}{(2 \pi)^3} V\left(-\frac{k^2}{\Lambda^2}\right) 
e^{i\vec{k} \cdot \vec{x}} \nonumber\\
&=&\frac{M}{2\pi^2} \int_0^\infty k^2 \frac{\mathrm{sin} kr}{kr} 
V\left(-\frac{k^2}{\Lambda^2}\right) \mathrm{d} k \, ,
\ee
where $r=\left| \vec{x} \right| $. This is the representation for effective energy 
density, and, once the form factor is specified, it can be numerically solved to get
$\rho^e(r)$. The radial integral contains the term $\frac{{\rm sin} kr}{kr}$. If we
expand the sine function, we have
\begin{equation}
\rho^e(r)=\frac{M}{2 \pi^2} \int_0^\infty k^2 V \left(-\frac{k^2}{\Lambda^2} \right) 
\mathrm{d}k+O(r^2) \, .
\end{equation}
Independently of the choice of the form factor $V(z)$, the leading term is a constant,
which means that at $r=0$ we will always have a positive effective energy density
proportional to the mass $M$ and fully determined by $V(z)$. As long as the 
convergence velocity of $V(z)$ is larger than $k^{-3}$, we are able to get a finite 
effective energy density. In classical general relativity, $V(z) = 1$ and the result
is not finite.

The covariant conservation and the additional condition $g_{tt}=-g_{rr}^{-1}$
completely determines $\mathcal{S}_{\mu\nu}$. 
The Einstein equations give
\be\label{einsteineq}
\frac{\mathrm{d}m}{\mathrm{d}r}&=& \frac{\kappa^2}{4} \rho^e r^2 \, , \\
\frac{1}{F} \frac{\mathrm{d}F}{\mathrm{d}r}&=&
\frac{2[m(r)+ \frac{\kappa^2}{4} p^e_r r^3]}{r[r-2m(r)]} \, , \\
\frac{\mathrm{d}p^e_r}{\mathrm{d}r}&=&-\frac{1}{2F}
\frac{\mathrm{d}F}{\mathrm{d}r}
(\rho^e+p^e_r)+\frac{2}{r}(p^e_\theta-p^e_r) \, .
\ee
From Eq.~(\ref{einsteineq}), we find the mass enclosed in the radius $r$
\begin{equation}
m(r)=\frac{\kappa^2}{4} 
\int_0^r \mathrm{d}r' r'^2 \rho^e(r') \, .
\end{equation}
At $r=0$, $m=0$ and $F=1$. For $r \to \infty$, $m$ is a constant and 
$F(r)\to1$. We have thus solutions with two or more horizons. Moreover,
at $r=0$ a constant energy density gives a de~Sitter spacetime, independently
of the choice of the form factor. 
Here the effective cosmological constant is of order $\kappa^2 M \Lambda^3$
and this, as argued at the end of this paper, may be the key-ingredient to
address our question.

We now consider two specific form factors, proposed respectively by 
Krasnikov~\cite{ff-k} and by Tomboulis~\cite{ff-t}:
\be\label{kras}
V(z) &=& e^{-|z|^n} \, , \\
V(z) &=& e^{-\frac{1}{2} [ \gamma_E +\Gamma (0, p_{\gamma+1} (z)^2) 
+ {\rm ln} \, p_{\gamma+1} (z)^2 ]} \, . \label{tomb}
\ee
Here $p_{\gamma+1}(z)$ is a polynomial of order $\gamma+1$. The 
super-renormalizability of the theory requires $\gamma\ge3$ and in what 
follows we will only consider the minimal renormalizable theory with 
$\gamma=3$. In the low energy limit, $z\equiv -\Box_\Lambda \to 0$, and to
recover general relativity we need $V(z)\to1$. So $p_{\gamma+1}(0)=0$. 
Moreover, we should expect deviations from general relativity when $z\neq0$, 
so for any $z>0$ we have $p_{\gamma+1} (z)\neq0$. This argument is in accordance with 
the restriction for $p_{\gamma+1}(z)$. We can therefore consider three cases: 
$p_{\gamma+1}(z)=z^4$, and $p_{\gamma+1}(z)=z^4\pm6 z^3+10z^2$. The 
latter two cases are taken as a generalization of the first one.

\begin{figure*}
\begin{center}
\includegraphics[width=7cm]{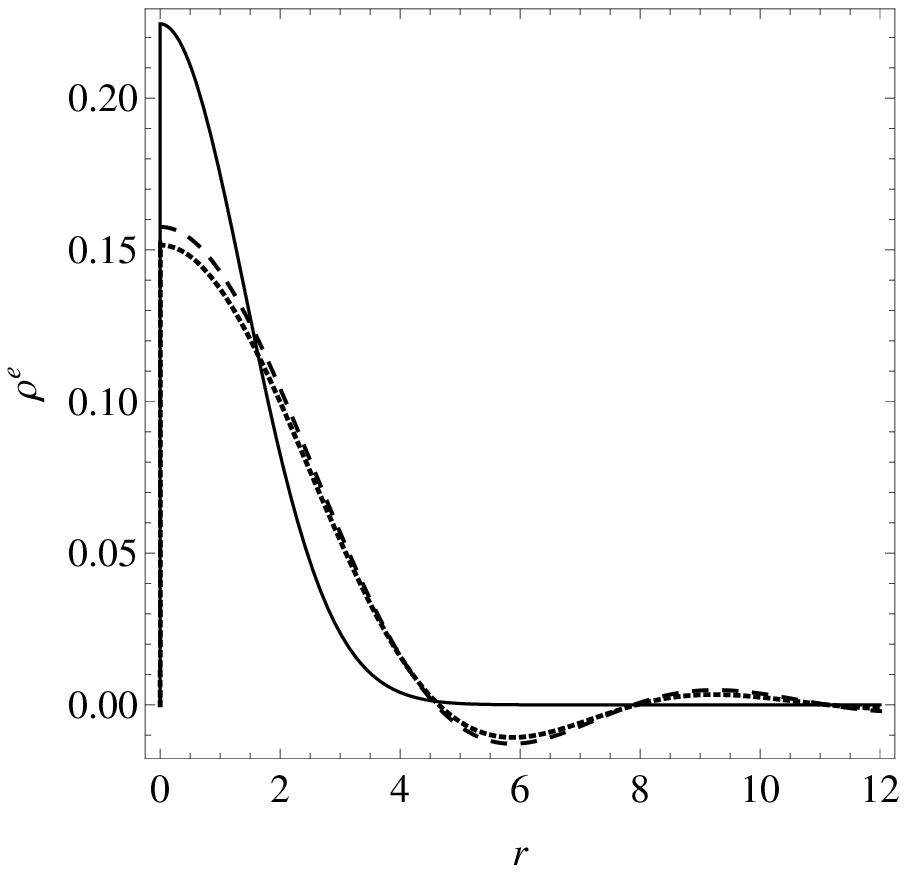}
\hspace{1cm}
\includegraphics[width=7cm]{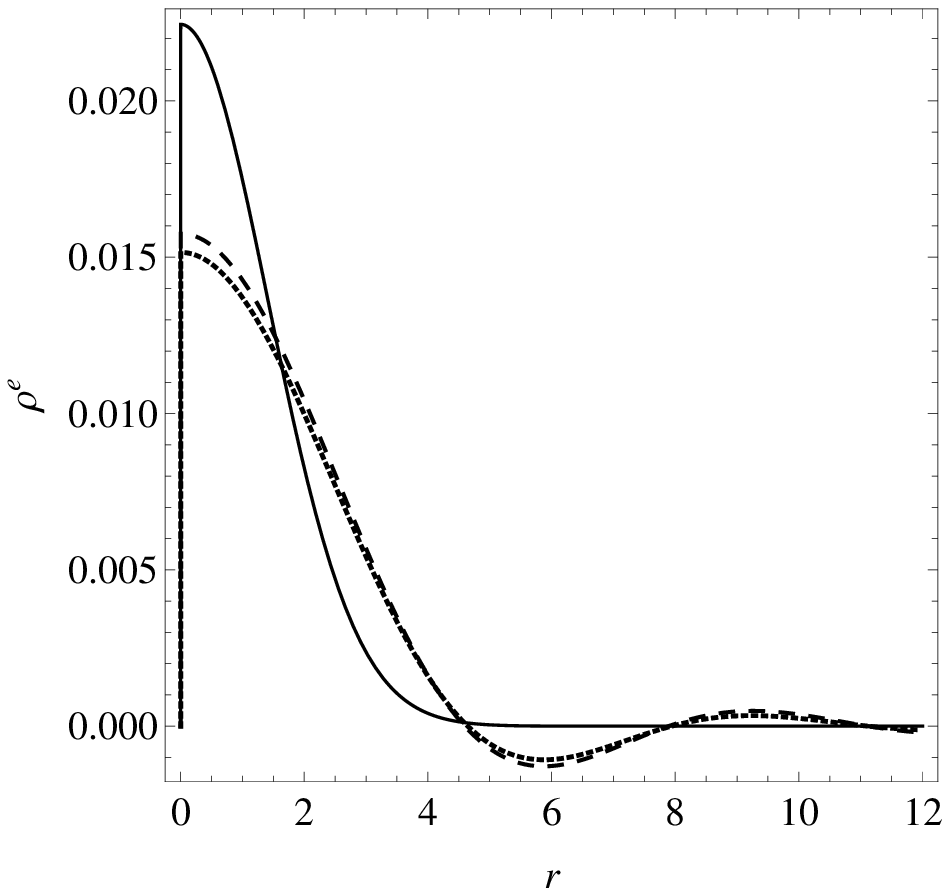}
\caption{Effective energy density as a function of the radial coordinate for the
Krasnikov form factor. Left panel: $\rho^e(r)$ for $M=10$ and $n=1$ (solid line),
5 (dotted line), and 10 (dashed line). Right panel: as in the left panel for $M=1$. 
The two plots have exactly the same shape and only differ in the scale of the 
vertical axis.}
\label{fig-BHKrasRho}
\vspace{1.5cm}
\includegraphics[width=7cm]{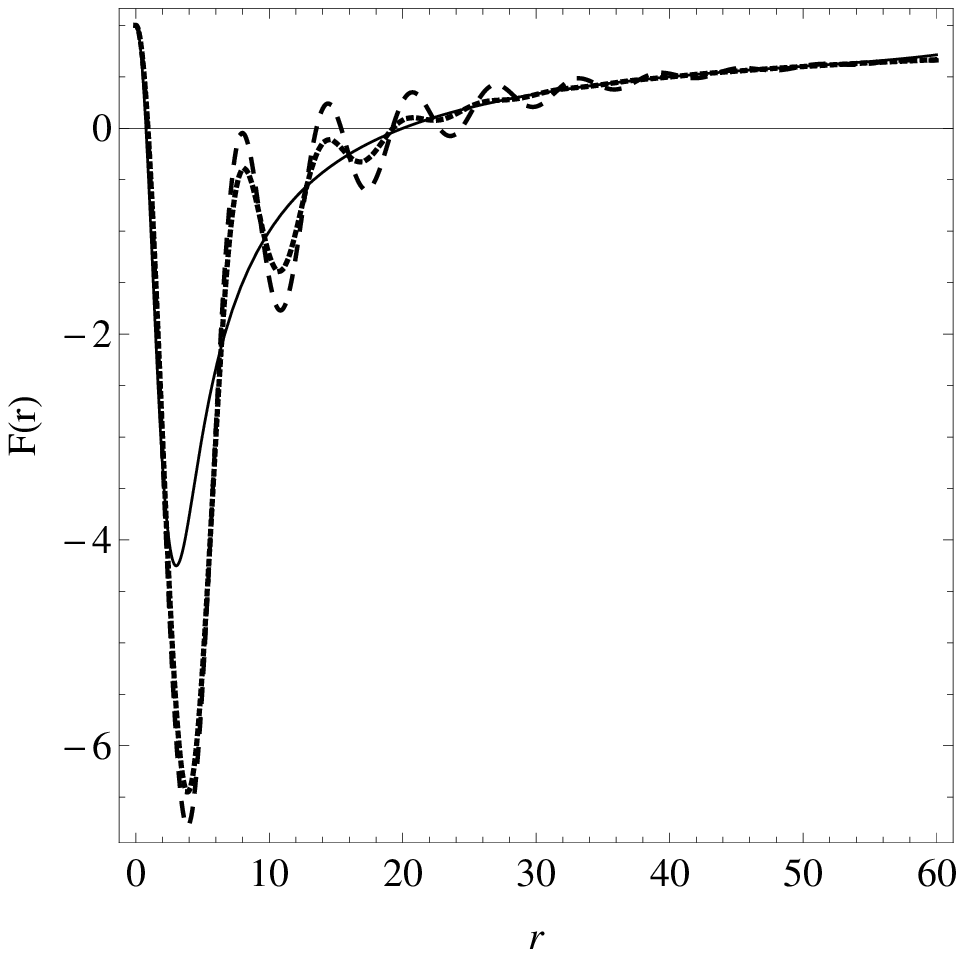}
\hspace{1cm}
\includegraphics[width=7cm]{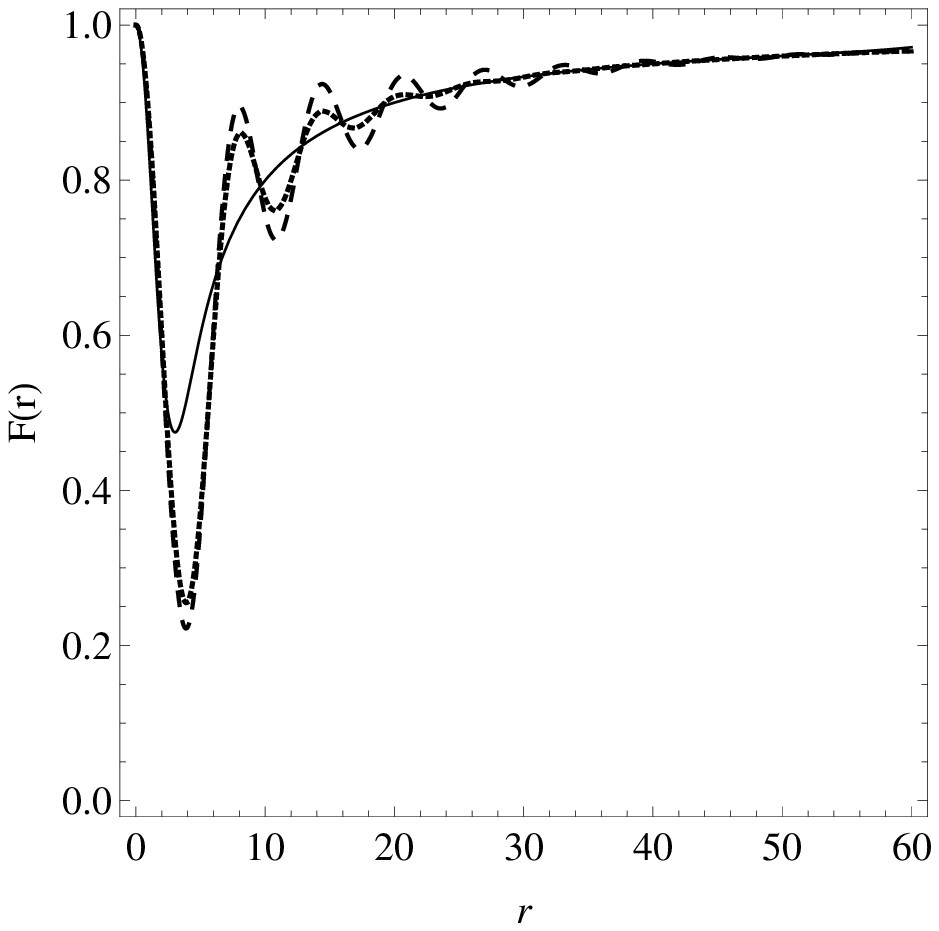}
\caption{$F$ as a function of the radial coordinate for the Krasnikov form factor.
Left panel: $F(r)$ for $M=10$ and $n=1$ (solid line), 5 (dotted line), and 10
(dashed line). The three curves cross at least two times the $F=0$ axis and
therefore every black hole has at least two horizons. Right panel: as in the
left panel for $M=1$. These configurations have no horizon has a consequence 
of the small mass.}
\label{fig-BHKrasFr}
\end{center}
\end{figure*}

Fig.~(\ref{fig-BHKrasRho}) shows $\rho^e(r)$ obtained from the numerical 
integration of Eq.~(\ref{eff-rho}) with the Krasnikov form factor and $n=1$, 5, 10. 
The left panel is for $M=10$, the right panel for $M=1$. Since $\rho^e(r) \propto M$,
if we change $M$ we change the scale, without altering the shape of $\rho^e(r)$.
For $n\ge 2$, $\rho^e$ is not monotonic and can assume negative values.
Because of that, it is possible to have more than two horizons. These plots show
also that the effective energy density $\rho^e$ approaches a finite value
for $r\to0$. Fig.~(\ref{fig-BHKrasFr}) shows the corresponding $F(r)$ functions. 
For $M=10$, we find at least two horizons, while for $M=1$ there is no horizons. 
For a large $n$, the oscillations of $\rho^e$ are stronger and therefore it is 
possible to form more than two horizons. However, this multi-horizon situation 
only exists when $M\sim10$ and $n$ is large. For instance, we found that there 
are just two horizons when $M=100$ and $n=10$.

\begin{figure*}
\begin{center}
\includegraphics[width=7cm]{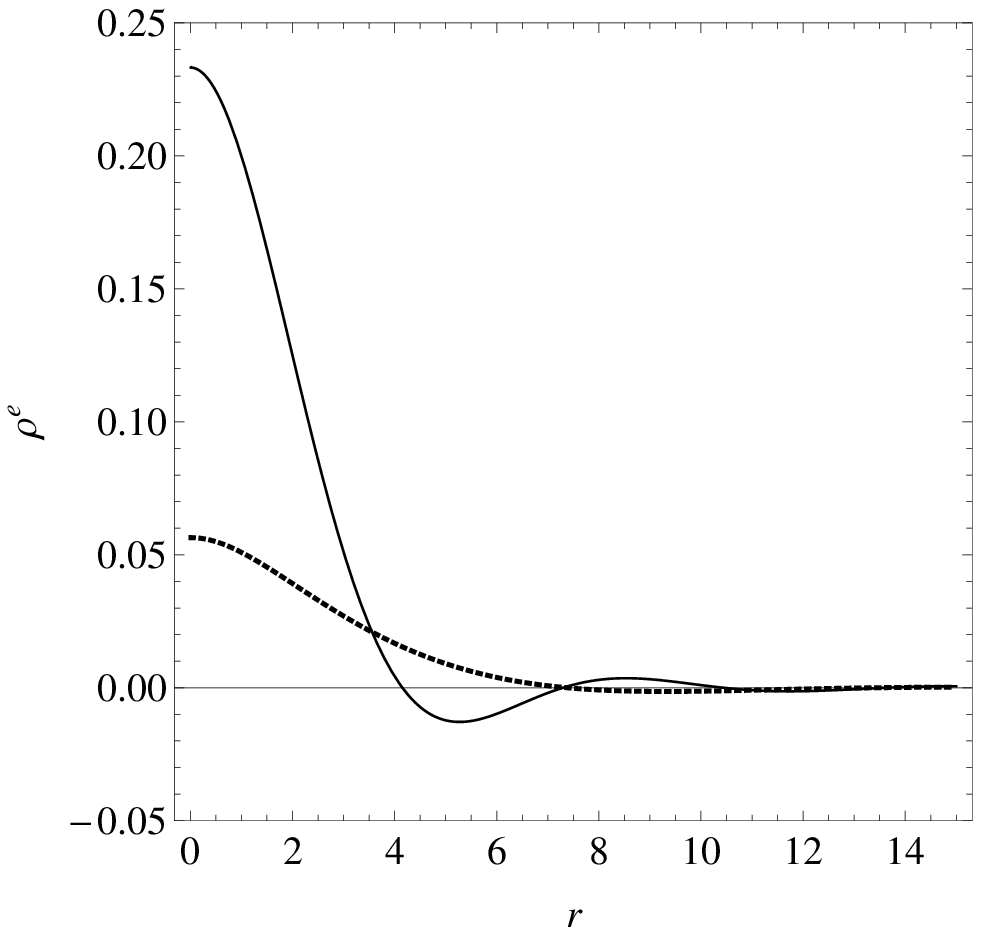}
\hspace{1cm}
\includegraphics[width=7cm]{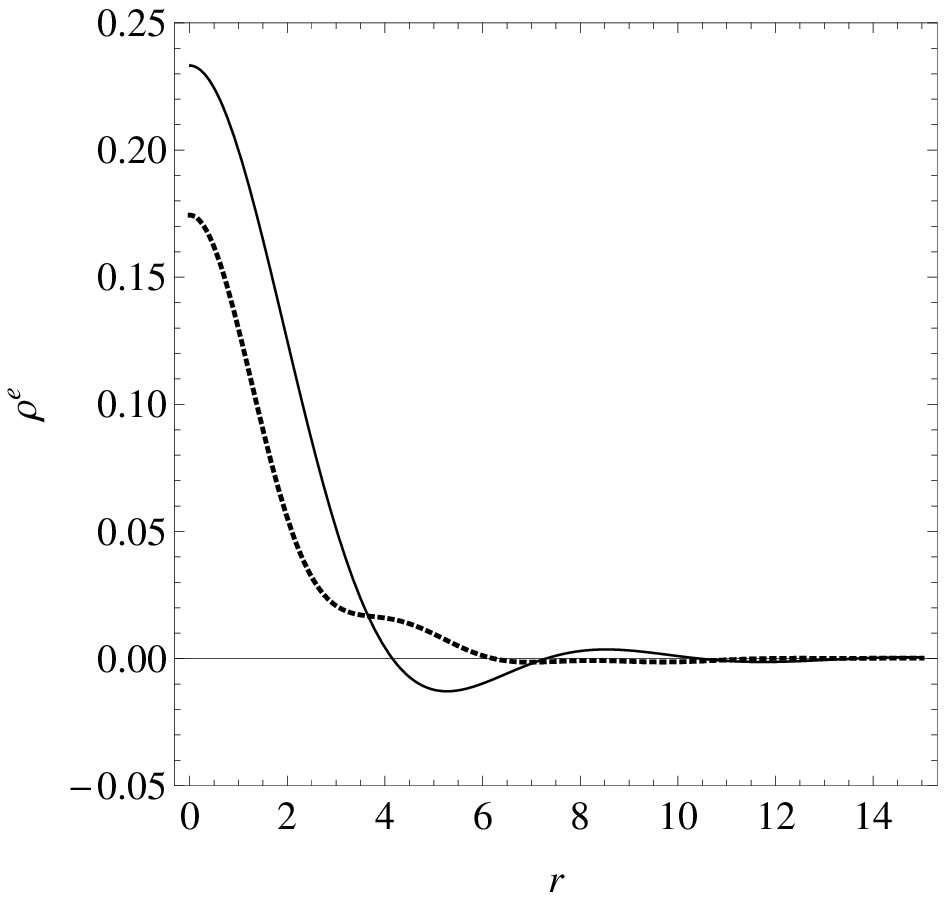}
\caption{Effective energy density as a function of the radial coordinate for the
Tomboulis form factor. Left panel: $\rho^e(r)$ for $M=10$ and $p_{\gamma+1}(z)=z^4$ 
(solid line) and $z^4+6z^3+10z^2$ (dashed line). Right panel: $\rho^e(r)$ for 
$M=10$ and $p_{\gamma+1}(z)=z^4$ (solid line) and $z^4-6z^3+10z^2$ 
(dashed line).}
\label{fig-BHTombRho}
\vspace{1.5cm}
\includegraphics[width=7cm]{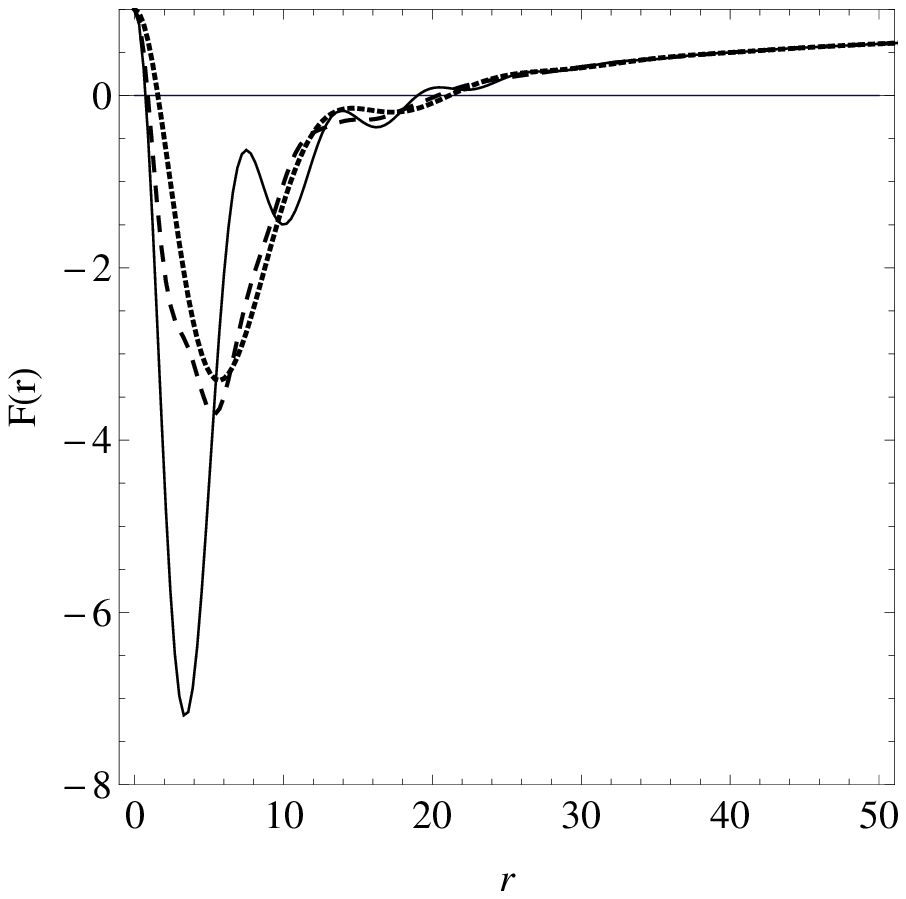}
\hspace{1cm}
\includegraphics[width=7cm]{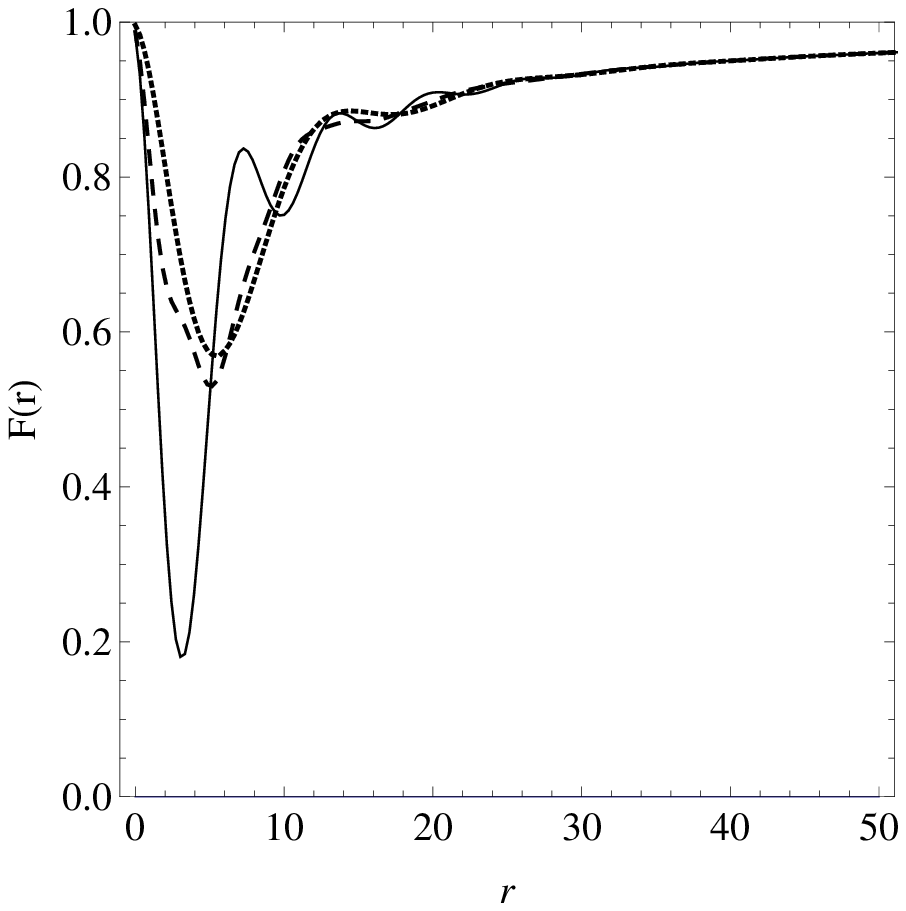}
\caption{$F$ as a function of the radial coordinate for the Tomboulis form factor.
Left panel: $F(r)$ for $M=10$ and $p_{\gamma+1}(z)=z^4$ (solid line), 
$z^4+6z^3+10z^2$ (dotted line), and $z^4-6z^3+10z^2$ (dashed line). The three 
curves cross two times the $F=0$ axis and therefore every black hole has two 
horizons. Right panel: as in the left panel for $M=1$. These configurations have 
no horizon has a consequence of the small mass.}
\label{fig-BHTombFr}
\end{center}
\end{figure*}

Figs.~\ref{fig-BHTombRho} and \ref{fig-BHTombFr} are for the Tomboulis form 
factor. Fig.~\ref{fig-BHTombRho} shows the effective energy density $\rho^e(r)$:
like for the Krasnikov form factor, $\rho^e$ is finite for $r=0$ and it has an
oscillatory behavior near $\rho^e = 0$, so that it can be negative at some radii.
Fig.~\ref{fig-BHTombFr} shows the function $F(r)$. For $M=10$ we have two
horizons, while there is no horizons when $M=1$. The choice of 
$p_{\gamma+1}(z)$ mainly affects the region $r < 2M$.

\section{Non-local gravity inspired collapse \label{s-4}}

Now we want to find the quantum-gravity corrected solution of the gravitational 
collapse for a homogeneous and spherically symmetric cloud. The scale factor 
$a(t)$ is determined through the propagator approach. We first write the metric 
as a flat Minkowski background plus a fluctuation $h_{\mu \nu}$,
\be 
g_{\mu \nu} &=& \eta_{\mu \nu} + \kappa \, h_{\mu \nu} \, , \nonumber\\
ds^2 &=& - dt^2 + a(t)^2 dx^i dx^j \delta_{i j} \, ,  
\ee
where $\eta_{\mu \nu} = {\rm diag}(-1,1,1,1)$. The conformal scale factor $a(t)$ 
and the fluctuation $h_{\mu\nu}(t, \vec{x})$ are related by the following 
relations:
\be
a^2(t) &=& 1 - \kappa h(t) \,  , \nonumber\\
h(t=-t_0) &=& 0 \, , \nonumber\\
 g_{\mu \nu}(t=-t_0) &=& \eta_{\mu\nu} \, , \nonumber\\
h_{\mu \nu}(t, \vec{x} ) &=& - h(t) \, {\rm diag}(0, \delta_{i j} ) 
\equiv - h(t) \, \mathcal{I}_{\mu \nu} \, .
\label{ah}
\ee
After a gauge transformation, we can rewrite the fluctuation in the usual harmonic 
gauge
\be
&& h_{\mu \nu}(x) \rightarrow h^{\prime}_{\mu \nu}(x) = 
h_{\mu \nu}(x)+ \partial_{\mu} \xi_{\nu} + \partial_{\nu} \xi_{\mu} \, , 
\nonumber\\
&& \xi_{\mu}(t) = \frac{3 \kappa}{2} \, 
\left( \int_0^{t} h(t') dt',0,0,0  \right) \, . 
\ee
The fluctuation now reads 
\be
&& h^{\prime}_{\mu \nu}(t, \vec{x} ) = 
h(t) \, {\rm diag} ( 3, - \delta_{i j} ) \, , \nonumber\\
&& h^{\prime \, \mu}_{\mu}(t, \vec{x} ) = - 6 h(t) .
\ee
We can then switch to the standard gravitational ``barred" field 
$\bar{h}^{\prime}_{\mu \nu}$ defined by 
\be
\bar{h}^{\prime}_{\mu \nu} = {h}^{\prime}_{\mu \nu} - \frac{1}{2} \eta_{\mu \nu} \, 
h^{\prime \, \lambda}_{\lambda} 
=  2 h(t) \, \mathcal{I}_{\mu \nu} \, ,
\ee
satisfying $\partial^{\mu} \bar{h}^{\prime}_{\mu \nu} = 0$.
The Fourier transform of $\bar{h}^{\prime}_{\mu \nu}$ is
\be
\tilde{\bar{h}}^{\prime}_{\mu \nu}(E, \vec{p}) 
= 2 \tilde{h}(E) (2 \pi)^3 \delta^3(\vec{p}) \, \mathcal{I}_{\mu \nu} \, . 
\label{FThxx}
\ee

The classical solution for the homogeneous and spherically symmetric 
gravitational collapse is known. We can thus compute the Fourier transform 
$\tilde{h}(E)$ defined in~(\ref{FThxx}). For $\omega \neq -1$, we have
\begin{equation}\label{htilde-general}
\tilde{h}(E)=\frac{2\pi\delta(E)}{\kappa}
+\frac{2\Gamma(\frac{4}{3(\omega+1)}+1) 
\mathrm{sin} (\frac{\pi}{2}\frac{4}{3(\omega+1)})}
{\kappa t_0^{\frac{4}{3(\omega+1)}} \left| E \right| ^{\frac{4}{3(\omega+1)}+1}} \, .
\end{equation}
In the case of radiation and dust, we have
\be
\tilde{h}(E)&=&\frac{2\pi\delta(E)}{\kappa}
+\frac{2}{\kappa t_0 E^2}, \;\;\; 
\text{(radiation)} \label{htilde-radi}\\
\tilde{h}(E)&=&\frac{2\pi\delta(E)}{\kappa}+
\frac{4\Gamma(\frac{4}{3})}{\sqrt{3} \kappa t_0^{4/3} \left| E \right| ^{7/3}}, 
\;\;\; \text{(dust)} \label{htilde-dust} \, .
\ee
The gauge independent part of the graviton propagator for the 
theory~(\ref{eq-theory}) and energy-momentum tensor $\tilde{T}^{\rho\sigma}(p)$ 
is~\cite{lll}
\be 
\hspace{-1cm}
&& \mathcal{O}^{-1}_{\mu \nu\rho\sigma}(p) = 
\frac{V(p^2/\Lambda^2) }{ 
p^2}   
\left( P^{(2)}_{\mu \nu\rho\sigma} - 
\frac{1}{2} P^{(0)}_{\mu \nu\rho\sigma}  \right) 
\nonumber\\
\hspace{-1cm}
&& \Longrightarrow \, 
\bar{h}^{\prime}_{\mu \nu}(x) = 
\kappa \! \int \! \frac{d^4 p}{(2 \pi)^4}
\mathcal{O}^{-1}_{\mu \nu\rho\sigma}(p) 
\tilde{T}^{\rho\sigma}(p) \, e^{i p x} \, ,
\label{propgauge2} 
\ee
where $P^{(2)}_{\mu \nu\rho\sigma}$ and 
$P^{(0)}_{\mu \nu\rho\sigma}$ are the graviton projectors,
\be
&& P^{(2)}_{\mu\nu\rho\sigma} =
\frac{1}{2} \left(\theta_{\mu\rho}\theta_{\nu\sigma} 
+ \theta_{\mu\sigma}\theta_{\nu\rho}\right) 
- \frac{1}{3} \theta_{\mu\nu}\theta_{\rho\sigma} \, , \nonumber\\
&& P^{(0)}_{\mu \nu\rho\sigma} =
\frac{1}{3} \theta_{\mu\nu}\theta_{\rho\sigma} \, , 
\ee 
and
$\theta_{\mu\nu} = \eta_{\mu\nu} - k_\mu k_\nu/k^2$.
Therefore
\be
&& h(t) = \kappa
\int \frac{d^4 p}{(2 \pi)^4} 
V \left(\frac{ - p^2}{\Lambda^2}\right) \, \tilde{h}(E, \vec{p} ) \,
e^{ i p x} = \nonumber\\
&& \hspace{0.7cm}
=  \kappa \int \frac{d E}{2 \pi} V \left(\frac{E^2}{\Lambda^2}\right) 
\, \tilde{h}(E) e^{- i E t} \, ,
\label{htg}
\ee
where
\be
\tilde{h}(E, \vec{p} ) = (2 \pi)^3 \delta^3(\vec{p}) \tilde{h}(E) \, .
\ee
If we know the classical solution $a(t)$, we can find the distribution
$\tilde{h}(E)$ that provides the correct solution for $V(-p^2/\Lambda^2) =1$.
We can then use a different form factor to find the corresponding quantum-gravity 
corrected scale factor $a(t)$. Here we want to find the solutions for the
gravitational collapse of a cloud of dust and radiation with the form factors
of Krasnikov and Tomboulis.

\begin{figure*}
\begin{center}
\includegraphics[width=7cm]{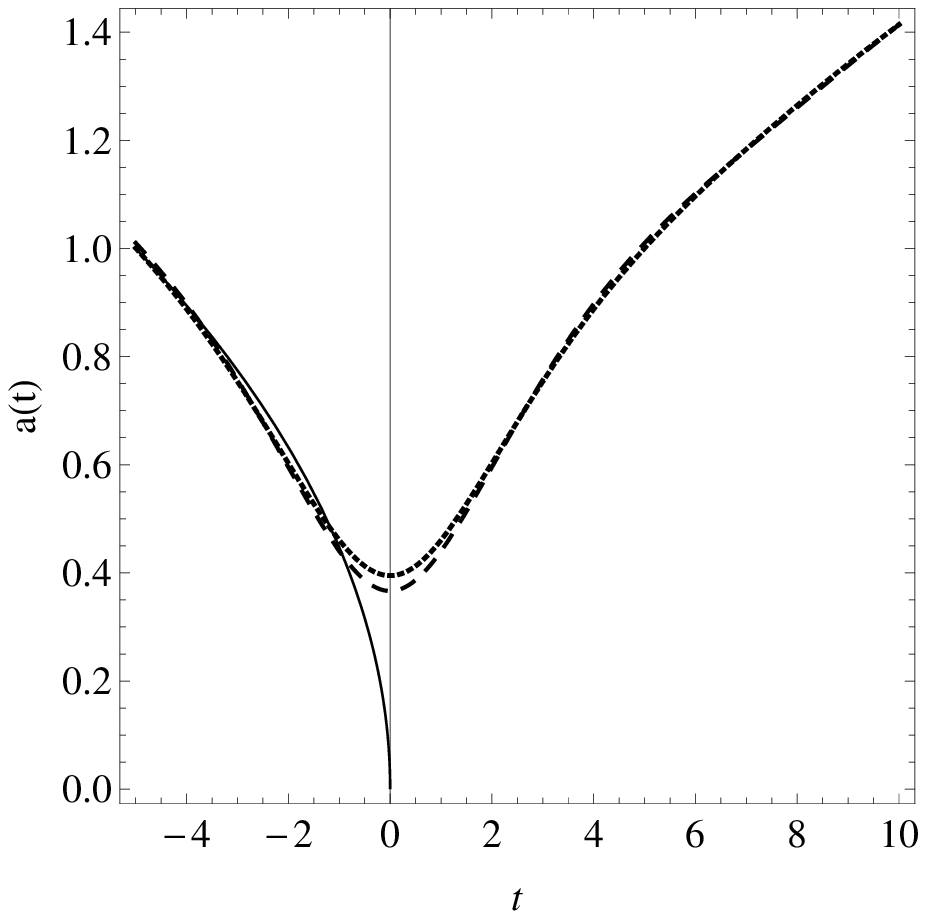}
\hspace{1cm}
\includegraphics[width=7cm]{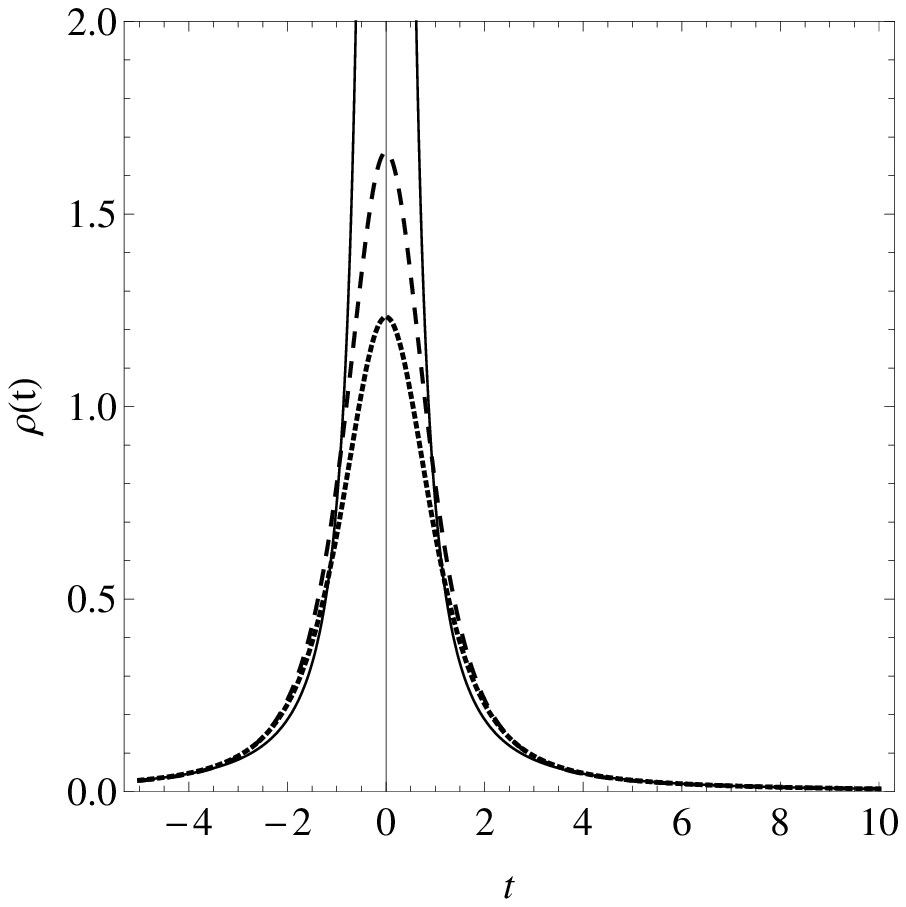}
\caption{Scale factor $a(t)$ (left panel) and energy density $\rho(t)$ (right panel) 
for the gravitational collapse of a homogeneous and spherically symmetric
cloud of radiation in the Krasnikov model. The solid line is for the classical
solution and the singularity occurs at the time $t=0$ when $a=0$. The dotted 
line is for the Krasnikov model with $n=2$, while the dashed line for the one
with $n=6$.}
\label{fig-ClpsKrasRadi}
\vspace{1.5cm}
\includegraphics[width=7cm]{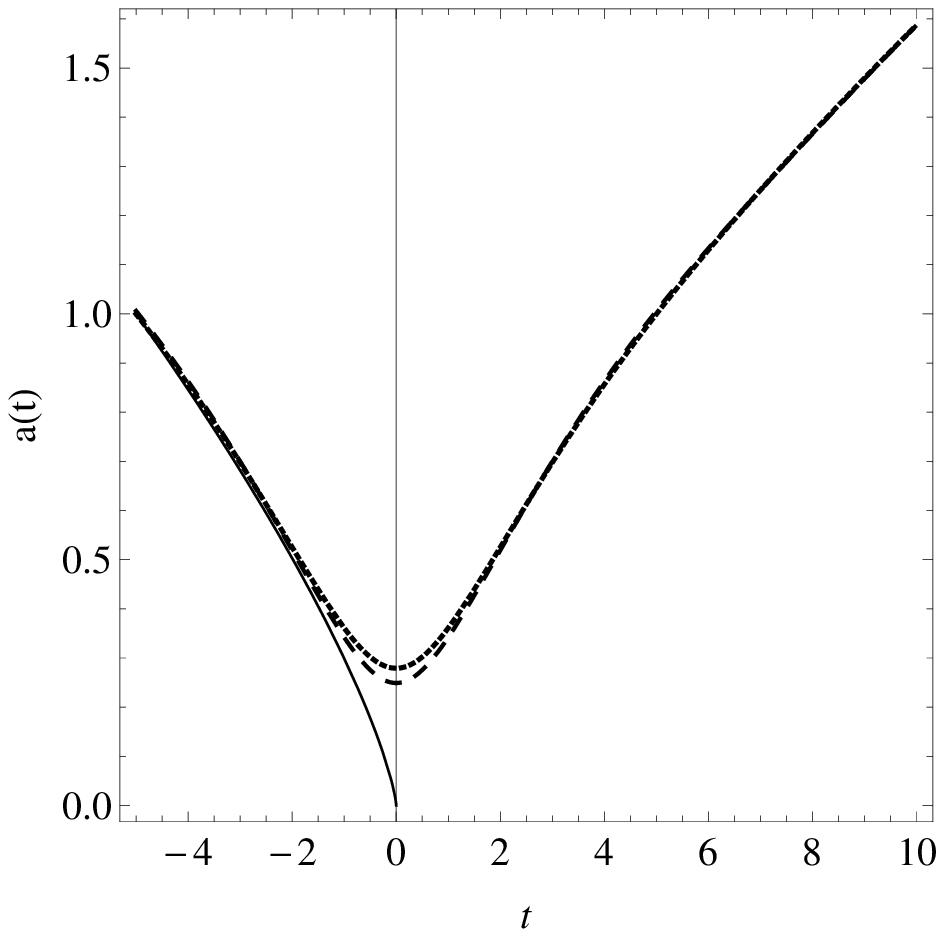}
\hspace{1cm}
\includegraphics[width=7cm]{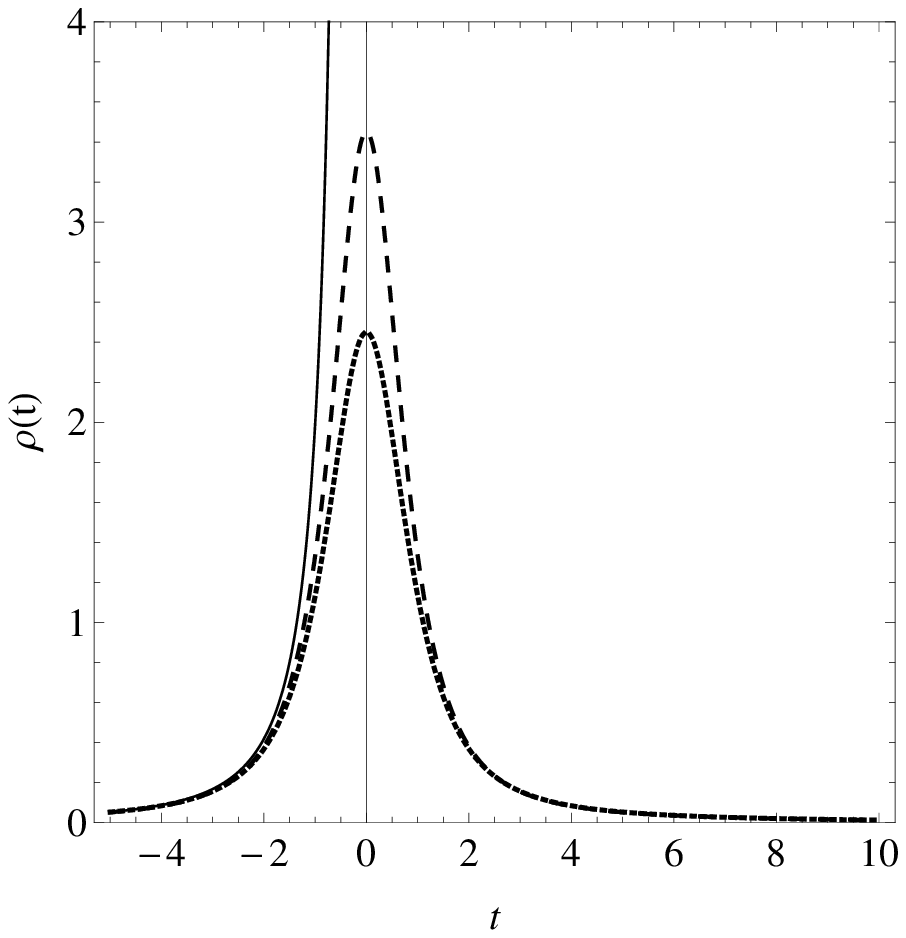}
\caption{As in Fig.~\ref{fig-ClpsKrasRadi} for the dust case.}
\label{fig-ClpsKrasDust}
\end{center}
\end{figure*}

\begin{figure}
\includegraphics[width=7cm]{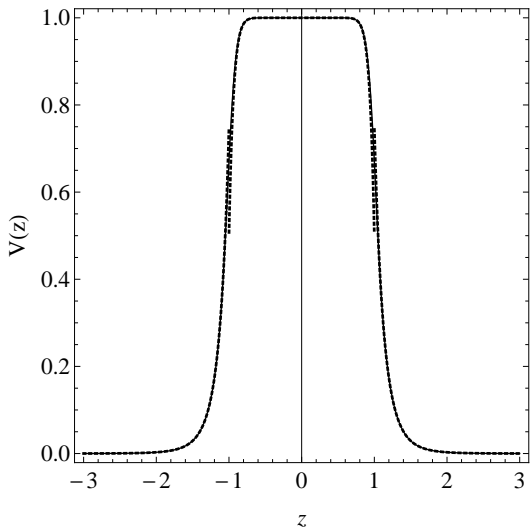}
\caption{Exact (solid line) and approximated (dashed line) Tomboulis form factor.}
\label{fig-TombApprox}
\end{figure}

\begin{figure*}
\begin{center}
\includegraphics[width=7cm]{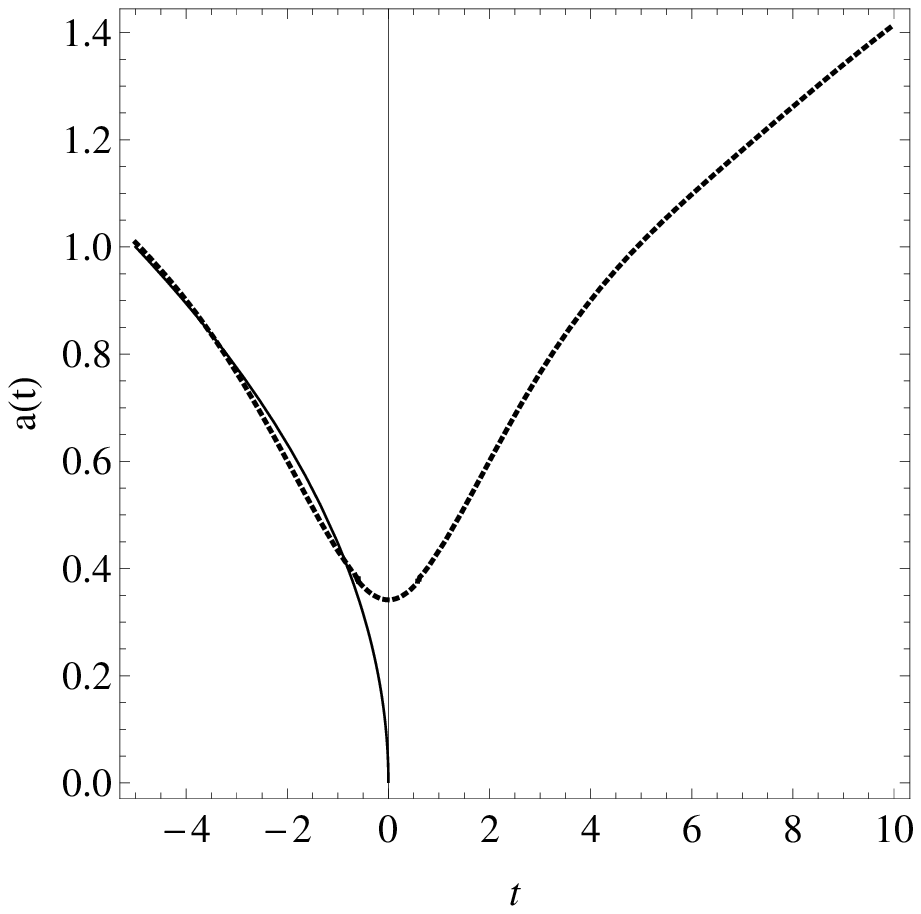}
\hspace{1cm}
\includegraphics[width=7cm]{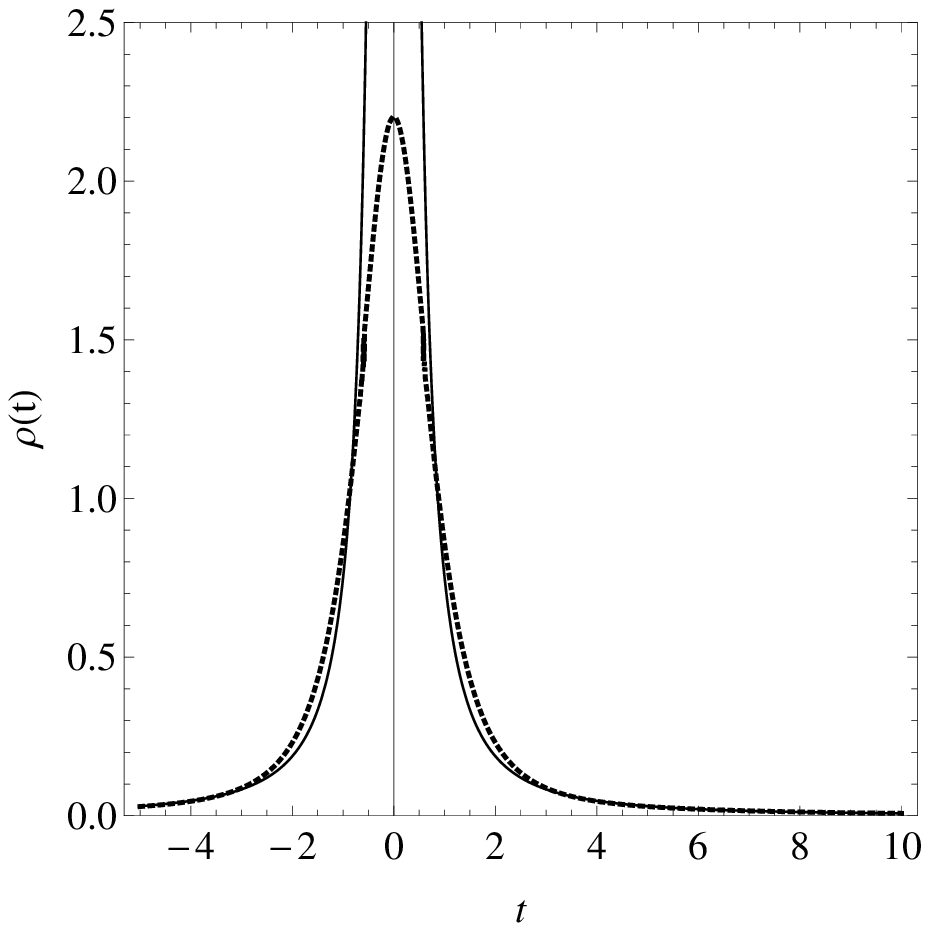}
\caption{Scale factor $a(t)$ (left panel) and energy density $\rho(t)$ (right panel) 
for the gravitational collapse of a homogeneous and spherically symmetric
cloud of radiation in the Tomboulis model. The solid line is for the classical
solution and the singularity occurs at the time $t=0$ when $a=0$. The dotted 
line is for the Tomboulis model with $p_{\gamma+1}(z)=z^4$.}
\label{fig-ClpsTombRadi}
\vspace{1.5cm}
\includegraphics[width=7cm]{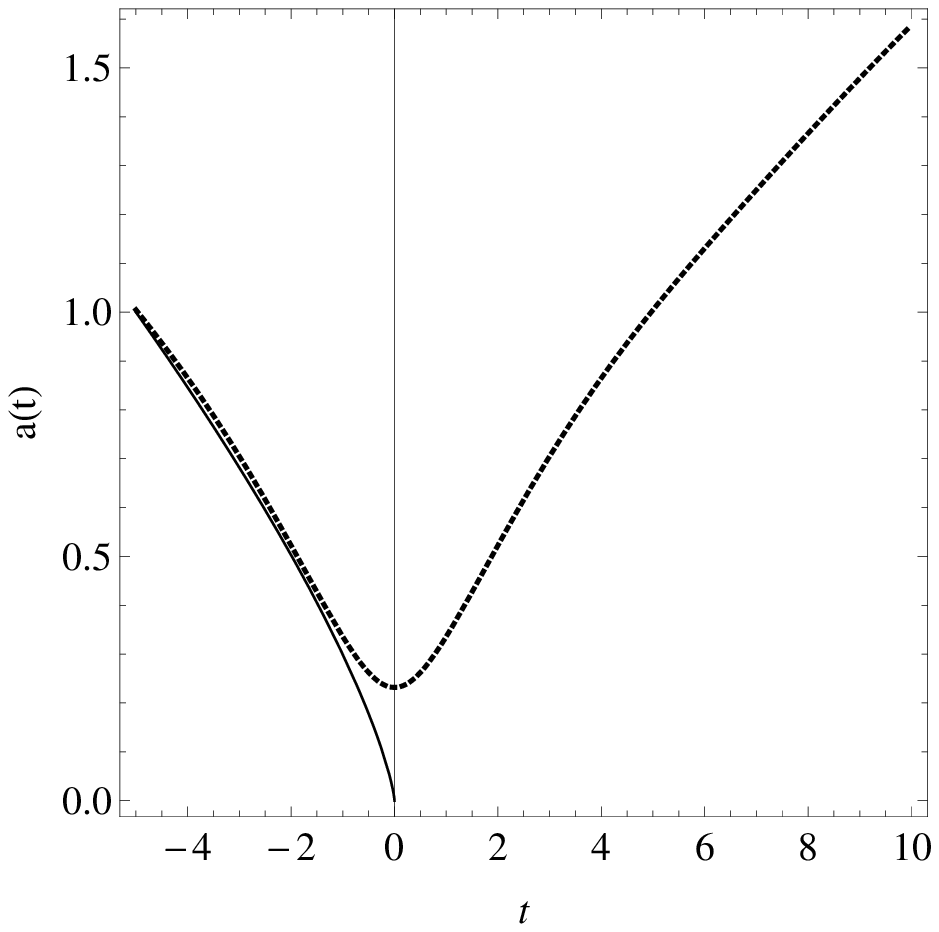}
\hspace{1cm}
\includegraphics[width=7cm]{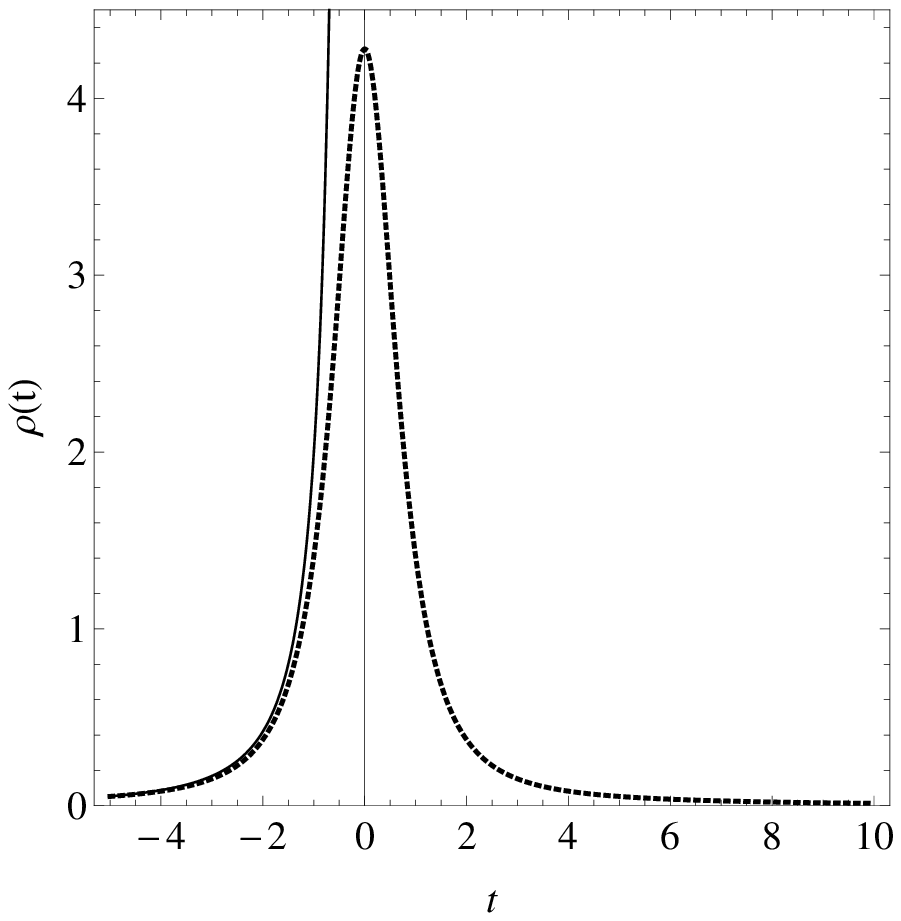}
\caption{As in Fig.~\ref{fig-ClpsTombRadi} for the dust case.}
\label{fig-ClpsTombDust}
\end{center}
\end{figure*}

Fig.~\ref{fig-ClpsKrasRadi} shows the scale factor $a(t)$ (left panel) and the energy
density (right panel) for the Krasnikov form factor with $n=2$ and 6 in the case of a 
cloud of radiation, $\omega = 1/3$. The dust case is shown in Fig.~\ref{fig-ClpsKrasDust}.
Classically, the scale factor $a(t)$ monotonically decreases and finally vanishes.
The corresponding energy density $\rho(t)$ therefore diverges at the time $t=0$.
In the quantum-gravity corrected picture, we find a bounce: $a(t)$ reaches a minimum
and then start increasing. Far from the bounce, the classical and the quantum solutions
are similar, while the difference becomes important at high densities.

The calculations with the Tomboulis form factor turn out to be significantly more
complicated. We thus use the following approximated form factor
\begin{equation}\label{eq-approx}
V(z)\approx
\begin{cases}
e^{-\frac{1}{2}p_{\gamma+1}(z)^2} \approx 1-\frac{1}{2}p_{\gamma+1}(z)^2 & \text{if } \left| p_{\gamma+1}(z) \right| \le 1 \\
e^{-\frac{\gamma_E}{2}}\frac{1}{\left| p_{\gamma+1}(z) \right| } & \text{if } \left| p_{\gamma+1}(z) \right| > 1
\end{cases}
\end{equation}
As shown in Fig.~(\ref{fig-TombApprox}), such an approximated form factor is
very similar to the exact one, with a tiny deviation near $z=\pm1$. While such an
approximated form factor has a discontinuity at $z=\pm1$, the Fourier transform 
is applicable as the limits at $z=\pm1$ in both directions are well-defined. Here
we have only studied the minimal renormalizable theory with $\gamma=3$ and 
have chosen $p_{\gamma+1}(z)=z^4$. The part $\left| p_{\gamma+1}(z) \right|> 1$ 
is trivial and we can obtain an analytic result. For the other part of the integral, 
we need to use a regularization prescription and separate the finite and the 
divergent parts. The detailed calculation is in Appendix~\ref{apxa}. Figs.~\ref{fig-ClpsTombRadi}
and \ref{fig-ClpsTombDust} show, respectively, the radiation and dust case.
The qualitative behavior of the scale factor $a(t)$ and the energy density $\rho(t)$ 
is the same as the one of the Krasnikov model.

\section{Summary and conclusions \label{s-5}}

In the present paper, we have studied both the static black hole solution and 
the homogeneous spherically symmetric collapse of a cloud of matter in a 
super-renormalizable and asymptotically free theory of gravity. The spacetime
singularity predicted in classical general relativity is removed in both the cases.
In the literature there were so far some scattered results in different theoretical
frameworks. Here we have studied this issue in more details within the Krasnikov
and Tomboulis models.

Static and spherically symmetric singularity free black hole solutions have been
obtained. At the origin, the effective energy density is always finite and positive,
independently of the exact expression of the form factor $V(z)$. In other words,
these black holes have a de~Sitter core in their interior,
where the effective cosmological constant is of order $\kappa^2 M \Lambda$, 
$\kappa^2 = 32\pi G_{\rm N}$, $M$ is the black hole mass, and $\Lambda$ 
is the energy scale of the theory which is naturally to expect to be close to the 
Planck mass. The singularity of the
spacetime is therefore avoided due to the repulsive behavior of the gravitational
force. For a large family of form factors, the effective energy density can be negative
in some regions, which eventually provides the possibility of having 
{\it multi-horizon black holes}. In the homogeneous and spherically symmetric 
collapse of a cloud of matter, the formation of the singularity is always replaced by a 
bounce. Far from the bounce, the collapse follows the classical solution, while it 
departs from it at high densities. Strictly speaking, asymptotic freedom is sufficient 
to remove the singularity, but the presence of a bounce requires also a repulsive 
character for gravitational field in the high energy regime.

In conclusion, we have provided some convincing examples that show how the
final products of the quantum-gravity corrected collapse solutions are not the
quantum-gravity corrected Schwarzschild black hole metrics. This is not the
result that one would expect {\it a priori}. 
There may be 3 natural explanations.
\begin{enumerate}
\item Static regular black holes cannot be created in any physical process. 
In this case, even if they are solution of a theory, they are much less interesting 
than their classical counterparts that can be created in a collapse.
\item The final product of the gravitational collapse is not unique. The collapse 
of a homogeneous and spherically symmetric cloud of matter does not produce
a static regular black hole, but the collapsing matter bounces and then expands. 
With different initial conditions, not known at the moment, static regular black 
holes may form.
\item The simple example of a homogeneous cloud of matter oversimplifies the
picture and misses important physics. As discussed in Section~\ref{s-2}, in the 
classical dust case we have a homogeneous interior and a Schwarzschild
exterior without ingoing or outgoing flux through any spherical shell of comoving
radial coordinate $r$. However, that is not true in general, and the exterior 
spacetime is a generalized Vaidya solutions with ingoing or outgoing flux of 
energy. This means that the homogeneous solution is not stable and must 
evolve to an inhomogeneous model. While the bounce can still occur, after it 
the collapsing matter may not expand forever. The boundary effects are 
important and, after proper readjustment that can unlikely be described without 
a numerical strategy, the collapse approaches the static black hole solution. 
\end{enumerate}
The possibility 1 excludes the possibilities 2 and 3, but the latter may also
coexist. Here we have focused on the asymptotically free
gravity theory with the Lagrangian given in Eq.~(\ref{eq-theory}), and we have 
shown that the issue indeed exists. The theoretical model is not sick, and 
therefore we cannot attribute the problem to the fact that we are considering 
a non-consistent quantum theory.
However, it is easy to compute the propagator
for this class of theories around de Sitter spacetime background and to show the presence of a ghost
when the cosmological constant exceeds the square of the Planck mass (in preparation).
This fact may explain the fact that our static black holes are not the final product 
of the gravitational collapse. These black holes have indeed a de~Sitter core in 
which the effective cosmological constant is $\kappa^2 M \Lambda$. If the black 
hole mass $M$ exceeds the Planck mass, there is a ghost and the black hole is 
unstable. Therefore the solutions here presented cannot be the final product,
but only an intermediate phase of the gravitational collapse.


\begin{widetext}

\begin{acknowledgments}
This work was supported by the NSFC grant No.~11305038, 
the Shanghai Municipal Education Commission grant for Innovative 
Programs No.~14ZZ001, the Thousand Young Talents Program, 
and Fudan University.
\end{acknowledgments}


\appendix

\section{Theoretical framework} \label{appendix-0}

For the theories studied in this paper, the classical action is
\be
\hspace{-0.2cm}
\mathcal{S} = \int d^4 x \frac{2 \sqrt{|g|}}{\kappa^2} 
\Big[R - G_{\mu\nu} \frac{ V(-\Box/\Lambda^2)^{-1} -1}{\Box} 
R^{\mu\nu} \Big] \, ,
\label{theory}
\ee
where $G_{\mu\nu}$ is the Einstein tensor and $\kappa^2 = 32 \pi G_{\rm N}$. 
All the non-polynomiality is in the form factor $V(-\Box/\Lambda^2)$, which
must be an entire function. $\Lambda$ is the Lorentz invariant energy scale
and it is not subject to infinite or finite (non analytic) renormalizations. The 
natural value of $\Lambda$ is of order the Planck mass and in this case all
the observational constraints are satisfied. Indeed, at classical level all the 
corrections to the Einstein-Hilbert action are suppressed by $1/\Lambda$, 
and if the value of $\Lambda$ is large the theory can reduce to general 
relativity at low energies. At quantum level, the introduction of non-local 
operators in the action could potentially lead to strong non-localities 
generated by the renormalization group flow towards the infrared, in 
disagreement with observations. This is not the case here, as a consequence 
of the Donoghue argument~\cite{donog} and of the fact that our action only 
involves entire functions.

The entire function $V(-\Box/\Lambda^2)$ must have no poles in the whole 
complex plane, in order to ensure unitarity, and must exhibit at least logarithmic 
behavior in the ultraviolet regime, to give super-renormalizablitity at the quantum 
level. The theory is uniquely specified once the form factor is fixed, because the 
latter does not receive any renormalization: the ultraviolet theory is dominated 
by the bare action (that is, counterterms are negligible). In this class of theories, 
we only have the graviton pole. Since $V(z)$ is an entire function, there are 
no ghosts and no tachyons, independently of the number of time derivatives 
present in the action. This is the main reason to introduce a non-polynomial Lagrangian.
At the phenomenological level the form factor could be 
experimentally constrained, for example measuring the corrections to the 
gravitational potential, or hypothetically measuring a cross section in a scattering 
process at high energy. 
Concerning the difficulties with particular form factors and non-local operators,
we note that the class of operators introduced by Krasnikov and by Tomboulis
are well defined in the Euclidean as well as in the Lorentzian case, because 
$(k_{\rm E}^2)^2 = (k^2)^2$, where $k_{\rm E}$ is the momentum in the Euclidean 
space~\cite{ff-k}.

More details on the ultraviolet and infrared properties of this class
of theories can be found in the last paper in~\cite{bounce}. We note that here it is 
possible to anti-screen gravity in the UV without introducing extra degrees of freedom because the theory is characterized by an entire function that goes to zero in the UV. We present below two different ways to see this.

{\it Tree-level unitarity ---}
The general and clear way to address the unitarity problem in Lagrangian formalism 
can be summarized as follows (for more details, see~\cite{v3-r}): 1) we calculate the 
propagator expanding the action to the second order in the graviton fluctuation, 
2) we calculate the amplitude with general external energy tensor sources, and 
3) we evaluate the residue at the poles. A general theory is well defined if  
``tachyons" and ``ghosts" are absent, in which case the corresponding propagator 
has only first poles at $k^2 - M^2 =0$ with real masses (no tachyons) and with 
positive residues (no ghosts). In our class of theories, we only have one pole 
in $k^2 = 0$ with positive residue.

When we introduce a general source, the linearized action including the 
gauge-fixing reads  
\be
\mathcal{L}_{hT} = \frac{1}{2} h^{\mu \nu} \mathcal{O}_{\mu \nu \rho \sigma} h^{\rho \sigma} 
- g \, h_{\mu \nu} T^{\mu \nu},
\label{LGM}
\ee
where
\be 
\hspace{-1cm}
\mathcal{O}^{-1}_{\mu \nu\rho\sigma}(p) = 
\frac{V(p^2/\Lambda^2) }{ 
p^2}   
\left( P^{(2)}_{\mu \nu\rho\sigma} - 
\frac{1}{2} P^{(0)}_{\mu \nu\rho\sigma}  \right) \, ,
\ee
$P^{(2)}_{\mu \nu\rho\sigma}$ and $P^{(0)}_{\mu \nu\rho\sigma}$ are the graviton projectors
\be
P^{(2)}_{\mu\nu\rho\sigma} =
\frac{1}{2} \left(\theta_{\mu\rho}\theta_{\nu\sigma} 
+ \theta_{\mu\sigma}\theta_{\nu\rho}\right) 
- \frac{1}{3} \theta_{\mu\nu}\theta_{\rho\sigma} \, , \quad
P^{(0)}_{\mu \nu\rho\sigma} =
\frac{1}{3} \theta_{\mu\nu}\theta_{\rho\sigma} \, , 
\ee 
and $\theta_{\mu\nu} = \eta_{\mu\nu} - k_\mu k_\nu/k^2$. The transition amplitude in momentum space is 
\be
\mathcal{A} = g^2 \, T^{\mu \nu} \, \mathcal{O}^{-1}_{\mu \nu \rho \sigma} \, T^{\mu \nu} \, ,
\label{ampli1}
\ee
where $g$ is an effective coupling constant. To make the analysis more explicit, we can expand the sources using the following set of independent vectors in the momentum space 
\be
k^{\mu} = (k^0, \vec{k}) \, , \,\, \tilde{k}^{\mu} = (k^0, - \vec{k}) \, , \,\,\,\, 
 \epsilon^{\mu}_i = (0, \vec{\epsilon}) \, , 
\,\, i =1, \dots , D-2 \, ,
\ee
where $\vec{\epsilon}_i$ are unit vectors orthogonal to each other and to $\vec{k}$. The symmetric stress-energy tensor reads 
\be
T^{\mu\nu} = a k^{\mu} k^{\nu} + b \tilde{k}^{\mu} \tilde{k}^{\nu} + c^{i j} \epsilon_i^{(\mu} \epsilon_j^{\nu)} + d \, k^{(\mu} \tilde{k}^{\nu)} + e^i k^{(\mu} \epsilon_i^{\nu)} + f^i \tilde{k}^{(\mu} \epsilon_i^{\nu)}.
\ee
The conditions $k_{\mu} T^{\mu \nu} =0$ and $k_{\mu} k_{\nu}T^{\mu \nu} =0$ place constrains 
on the coefficients $a,b,d, e^i, f^i$. 
Introducing the spin-projectors 
and the conservation of the stress-energy tensor  $k_{\mu} T^{\mu \nu} = 0$ in (\ref{ampli1}), the amplitude results 
\be
\mathcal{A} = g^2 \left\{  T_{\mu \nu} T^{\mu \nu} - \frac{T^2}{D-2} \right\} \frac{e^{- H(k^2/\Lambda^2)}}{k^2} \, ,
\label{ampli2}
\ee
where $T = T^\mu_\mu$. Clearly, there is only the graviton pole in $k^2 =0$ and the residue at $k^2=0$ is 
\be
{\rm Res} \, \left( \mathcal{A} \right) \big|_{k^2 =0} = g^2 \left[ (c^{ij})^2 - \frac{ (c^{ii})^2}{D-2} \right] \Big|_{k^2 =0}.
\label{residuo}
\ee
For $D>3$, we find that ${\rm Res} \, \left( \mathcal{A} \right) \big|_{k^2 =0}>0$ (because $H(0) =0$), which means that the theory is unitary. Instead, for $D=3$ the graviton is not a dynamical degree of freedom and the amplitude is zero.

As an example of this quantum transition, we can consider the interaction of two static point particles. Here $T^{\mu}_{ \nu} = {\rm diag}(\rho,0,0,0)$ with $\rho = M \, \delta(\vec{x})$ and the amplitude (\ref{ampli2}) simplifies to 
\be
\mathcal{A} = g^2 \, M^2 \left( \frac{D-3}{D-2} \right) \, \frac{e^{- H(k^2/\Lambda^2)}}{k^2} \, ,
\label{ampli3}
\ee
which is positive in $D>3$ and zero for $D=3$ since, again, there are no local degrees of freedom in $D=3$.

{\it K$\ddot{\rm a}$ll\'en-Lehmann (KL) representation ---}  
Any action in the class of theories here presented is defined in terms of an entire function with no zeros, singularities, nether poles. 
Therefore the usual derivation of the KL decomposition goes straightforward. 
Let us construct step by step such representation for a weakly nonlocal prototype theory. 
It is easy to show that,
\be
\int \frac{d^3 \vec{p} }{(2 \pi)^3 2 E_{\vec{p}}} = \int \frac{d^4 p }{(2 \pi)^4 } \delta(p^2 - m_{\lambda}^2) \theta(p^0) e^{- H(p^2 - m_{\lambda}^2 )} \, .
\ee
Therefore a complete set of states satisfy,
\be
{\bf 1} = | 0 \rangle \langle 0 | + \sum_{\lambda} \int \frac{d^3 \vec{p} }{(2 \pi)^3 2 E_{\vec{p}}} 
| \lambda_{\vec{p}} \rangle \langle  \lambda_{\vec{p}} | 
\ee
where we defined the state $\lambda_{\vec{p}}$ by boosting of an arbitrary boost $\vec{p}$ 
a momentum zero state $\lambda_0$ and we summed over all $\vec{p}$. 
Assuming that $\langle 0 | \phi(x) | 0 \rangle =0$ and Lorentz invariance of $\phi(0)$, 
\be
\langle 0 | \phi(x) | \lambda_{\vec{p}} \rangle = e^{- i p x} \langle 0 | \phi(0) | \lambda_{\vec{p}} \rangle 
= e^{- i p x} \langle 0 | \phi(0) | \lambda_{0} \rangle \Big|_{p_0 = E_{\vec{p}} }\, .
\ee
We thus get for $x_0 > y_0$:
\be
&& \langle 0 | \phi(x) \phi(y) | 0 \rangle \Big|_{x_0 > y_0} = 
\sum_{\lambda}  \int \frac{d^3 \vec{p} }{(2 \pi)^3 2 E_{\vec{p}}} e^{- i p (x-y)}  \Big|_{p_0 = E_{\vec{p}} } \, 
| \langle 0 | \phi(0) | \lambda_{0} \rangle |^2 \nonumber \\
&& \hspace{3.2cm} 
= \sum_{\lambda}  \int_{C_F} \frac{d^4 p }{(2 \pi)^4 } \frac{i }{(p^2 - m_{\lambda}^2) \, e^{H(p^2 - m_{\lambda}^2)}  } \, 
e^{- i p (x-y)} \, 
| \langle 0 | \phi(0) | \lambda_{0} \rangle |^2 \nonumber \\
&&\hspace{3.2cm} 
 = \sum_{\lambda}  e^{- H(- \Box - m_{\lambda}^2)}  \int_{C_F} \frac{d^4 p }{(2 \pi)^4 } \frac{i }{(p^2 - m_{\lambda}^2) \, } \, e^{- i p (x-y)} \, 
| \langle 0 | \phi(0) | \lambda_{0} \rangle |^2 \nonumber \\
&&\hspace{3.2cm} 
 \rar \sum_{\lambda}  e^{- H(- \Box - m_{\lambda}^2)}  \int_{  R } \frac{d^4 p }{(2 \pi)^4 } \frac{i }{(p^2 - m_{\lambda}^2 + i \epsilon ) \, } \, e^{- i p (x-y)} \, 
| \langle 0 | \phi(0) | \lambda_{0} \rangle |^2 \nonumber \\
&&\hspace{3.2cm} 
 = \sum_{\lambda}   \int_{  R } \frac{d^4 p }{(2 \pi)^4 } \frac{i \,   e^{- H(p^2 - m_{\lambda}^2)} }{(p^2 - m_{\lambda}^2 + i \epsilon ) \, } \, e^{- i p (x-y)} \, 
| \langle 0 | \phi(0) | \lambda_{0} \rangle |^2 ,
\ee
where the contour $C_F$ is closed in the lower half $p_0$-plane, while the integral along the real axes is 
evaluated displacing the poles for ${\rm Re}(p_0) > 0$ in the lower half plane and for 
${\rm Re}(p_0) < 0$ in the upper half plane (in our gravitational theory $m_{\lambda} = 0$).
Analog relation is obtained for $x_0<y_0$, therefore:
\be
\langle 0 | {\rm T} ( \phi(x) \phi(y) ) | 0 \rangle = \sum_{\lambda}   \int_{  R } \frac{d^4 p }{(2 \pi)^4 } \frac{i \,  e^{- H(p^2 - m_{\lambda}^2)} }{(p^2 - m_{\lambda}^2 + i \epsilon ) \, } \, e^{- i p (x-y)}  \, 
| \langle 0 | \phi(0) | \lambda_{0} \rangle |^2 \, .
\label{T}
\ee
The matrix element is Lorentz invariant and only depends on the mass of the state $m_{\lambda}$,
therefore we can write (\ref{T}) making use of the definite positive spectral density function $\rho(s)$ defined by,
\be
\rho(s) = \sum_{\lambda}  \delta(s - m_{\lambda}^2) \, 
| \langle 0 | \phi(0) | \lambda_{0} \rangle |^2 \, ,
\ee
as follows 
\be
\langle 0 | {\rm T} ( \phi(x) \phi(y) ) | 0 \rangle = \int_0^{+ \infty} \!\!\! ds \, \Delta_F(x-y;s) \rho(s) ,
\ee
where we defined the function 
\be
\Delta_F(x-y; m_{\lambda}^2)  = \int_{  R } \frac{d^4 p }{(2 \pi)^4 } \frac{i \,  e^{- H(p^2 - m_{\lambda}^2)} }{(p^2 - m_{\lambda}^2 + i \epsilon ) \, } \, 
e^{- i p (x-y)} \, , 
\ee
in analogy with the Feynman propagator. 
The spectral representation sees all the poles, but in this case they are the same as in general relativity, because we have introduced a function with no poles or zeros. In the case of a local theory with a finite number of derivatives, this is impossible and the theory is sick.

We expect  a single particle state to contribute to $\rho(s)$ as an isolated delta
\be
\langle 0 | {\rm T} ( \phi(x) \phi(y) ) | 0 \rangle = \int_0^{+ \infty} \!\!\! ds  \int_{  R } \frac{d^4 p }{(2 \pi)^4 } \frac{i \,  e^{- H(p^2 - m_{\lambda}^2)} }{(p^2 - m_{\lambda}^2 + i \epsilon ) \, } \, 
e^{- i p (x-y)} \, \delta(s - m_{\lambda}^2) \, Z \,\,\,\,\,\,\,\, ( Z>0) \, .
\ee

\section{Integrals with the Tomboulis form factor}\label{apxa}

We use the approximated form factor in Eq.~(\ref{eq-approx}). Let us first
consider the radiation case. For $\left| p_{\gamma+1}(z) \right| \ge 1$, the 
integral is trivial. When $\left| p_{\gamma+1}(z) \right| \le 1$, we have
\begin{equation}
\int_{-1}^{1} \frac{\mathrm{d}E}{2\pi} 
\Big(1-\frac{ p_{\gamma+1}(z)^2}{2}\Big) \frac{2}{t_0 E^2} 
e^{iEt} \, .
\end{equation}
The second term with $p_{\gamma+1}(z)^2$ is trivial, in the sense that it has an 
analytic result. We thus want to find the solution of the first term which, apart
a constant coefficient, has the form
\begin{equation}
I\equiv \int_{-1}^{1} \mathrm{d}E \frac{e^{\mathrm{i}Et}}{E^n} \, ,
\end{equation}
where $n$ is an even integer, so that the integral gives a real result and is 
an even function of $t$. In the case of radiation, $n=2$.

This integral can be calculated with the residue theorem. We introduce a small 
number $\mu$, so that the pole of order $n$ at $E=0$ is reduced to $n$ simple 
poles. To include all the poles, we can consider contour integrals of both the
upper and the lower circles. Writing $E=\rho e^{\mathrm{i}\theta}$,
$\mathrm{d}E=iE\mathrm{d}\theta$, and we have
\be
I+\int_{\substack{\theta=0 \\ \rho=1}}^{\theta=\pi}  \mathrm{d}E 
\frac{e^{\mathrm{i}Et}}{E^n} \equiv I+I_1 
=2\pi i\sum_{\substack{
\mu_i \text{in upper} \\ \text{unit circle} } } \mathrm{Res} 
\Big( \frac{e^{\mathrm{i}Et}}{E^n+\mu^n} \Big) 
+\pi i\sum_{\substack{
\mu_i \text{on} \\ \text{real axis} } } \mathrm{Res} 
\Big( \frac{e^{\mathrm{i}Et}}{E^n+\mu^n} \Big)
\ee
and
\be
I+\int_{\substack{\theta=0 \\ \rho=1}}^{\theta=-\pi}  
\mathrm{d}E \frac{e^{\mathrm{i}Et}}{E^n} \equiv I+I_2 
=2\pi i\sum_{\substack{
\mu_i \text{in the lower} \\ \text{unit circle} } } \mathrm{Res} 
\Big( \frac{e^{\mathrm{i}Et}}{E^n+\mu^n} \Big) 
-\pi i\sum_{\substack{
\mu_i \text{on the} \\ \text{real axis} } } \mathrm{Res} 
\Big( \frac{e^{\mathrm{i}Et}}{E^n+\mu^n} \Big) \, .
\ee
The combination of the two equations gives the result of $I$
\be
I=-\frac{1}{2} (I_1+I_2)+D \, ,
\ee
where
\be
I_1+I_2=\int_{\theta=0}^{\pi} \mathrm{d} \theta 
\Big\{ e^{-t\sin \theta} \sin [(n-1)\theta-t\cos \theta] 
+e^{t\sin\theta} \sin[(n-1)\theta+t \cos\theta] \Big\} \, ,
\ee
and the divergence $D$ of order $\mu^{n-1}$ is given by
\be
D=\pi i(\sum_{\substack{
\mu_i \text{in the upper} \\ \text{unit circle} } } -\sum_{\substack{
\mu_i \text{in the lower} \\ \text{unit circle} } } )\mathrm{Res} 
\Big( \frac{e^{\mathrm{i}Et}}{E^n+\mu^n} \Big)  \, .
\ee
For radiation, $n=2$ and we have
\be
I=-\frac{1}{2}\int_{\theta=0}^{\pi} \mathrm{d} \theta 
\Big\{ e^{-t\sin \theta} \sin [\theta-t\cos \theta]
+e^{t\sin\theta} \sin[\theta+t \cos\theta] \Big\} +\frac{\pi}{\mu} \, ,
\ee
where the divergent part is $D=\frac{\pi}{\mu}$ for $\mu\to0$. 

The dust case is
slightly different and it is more convenient to choose the exponential form for 
approximation of $V(z)$, so the integral is
\be
I \equiv \int_{-1}^{1} \mathrm{d}E \frac{1}{\left| E \right| ^\alpha} 
e^{-\frac{E^{16}}{2}} e^{\mathrm{i}Et} = 2 ~\mathrm{Re} \int_{0}^{1} 
\mathrm{d}E \frac{1}{ E ^\alpha} e^{-\frac{E^{16}}{2}} e^{\mathrm{i}Et} \, ,
\ee
where $\alpha>0$ is not an integer. With the transform for the integration variable
$E=\rho e^{i\theta}$, $\mathrm{d}E=\mathrm{i}E\mathrm{d}\theta$, 
and applying the contour integral we have 
\be
I_1+I_2+I_R+I_r \equiv 
\Big( \int_{\substack{\rho=0 \\ \theta=0}}^{\rho=1} + 
\int_{\substack{\rho=1 \\ \theta=2\pi}}^{\rho=0}
+\int_{\substack{\theta=0 \\ \rho=1}}^{\theta=2\pi} 
+\int_{\substack{\theta=2\pi \\ \rho\to0}}^{\theta=0} \Big)
 \mathrm{d}E \frac{1}{ E ^\alpha} e^{-\frac{E^{16}}{2}} e^{\mathrm{i}Et}=0 \, .
\ee
The first two terms $I_1+I_2$ can be simplified as
\be
I_1+I_2=2\int_0^1 \mathrm{d}\rho \frac{1}{\rho^\alpha} 
e^{-\frac{1}{2}\rho^{16}} e^{\mathrm{i}\rho t}
 e^{-\mathrm{i}\pi\alpha} (i\sin \alpha\pi) \, .
\ee
The integration of the unit circle gives the finite part of the required 
integral, and the integration around the origin gives the divergent 
part of order $\alpha-1$,
\be
I_R &=&i~\int_0^{2\pi} \mathrm{d}\theta 
~e^{-\frac{1}{2}\cos 16\theta-t\sin \theta} e^{i
[ -\frac{1}{2} \sin 16\theta +t \cos \theta + (1-\alpha) \theta ]} \, , \\
I_r &=&i~\int_{2\pi}^{0} \mathrm{d}\theta ~\frac{1}{\mu^{\alpha-1}} 
e^{-\frac{\mu^{16}}{2}\cos 16\theta-\mu t\sin \theta}
 e^{i[ -\frac{\mu^{16}}{2} \sin 16\theta +
 \mu t \cos \theta + (1-\alpha) \theta ]} 
 \simeq i~\int_{2\pi}^{0} 
 \mathrm{d}\theta ~\frac{e^{i (1-\alpha)\theta}}{\mu^{\alpha-1}} \, .
\ee
Finally, we can get the result
\be
I=-\frac{1}{\sin \alpha\pi} \int_0^{2\pi} \mathrm{d}\theta 
e^{-(\frac{1}{2}\cos 16\theta +t \sin \theta)} 
\cos [ -\frac{1}{2}\cos 16\theta +t \cos \theta 
+(1-\alpha)\theta +\alpha \pi ]
+\frac{2}{(\alpha-1)\mu^{\alpha-1}} \, .
\ee
For dust case, we have $\alpha=7/3$ and the divergence is of order $4/3$.

\end{widetext}


\end{document}